\documentclass{article}
\usepackage{hyperref}
\usepackage[utf8]{inputenc}
\usepackage{url}
\usepackage{graphicx}
\usepackage{multicol}        
\usepackage[bottom]{footmisc}
\usepackage[utf8]{inputenc}
\usepackage{algorithm}
\usepackage[noend]{algpseudocode}
\usepackage{array}
\usepackage{booktabs}
\usepackage{bold-extra}
\usepackage{color}

\usepackage{subcaption}
\usepackage{geometry}
\title{Exploratory Analysis of Academic Collaborations between French and US }

\author{George Panagopoulos, Michalis Vazirgiannis\\ {george.panagopoulos,mvazirg}@polytechnique.edu\\LIX, Ecole Polytechnique\\
Office for Science and Technology of the Embassy of France in the US}

\date{January 2021}

\begin{document}

\maketitle

\section{Introduction}

International academic collaborations cultivate diversity in the research landscape and facilitate multiperspective methods, as the scope of each country's science depends on its needs, history, wealth etc. Moreover the quality of science differ significantly amongst nations\cite{king2004scientific}, which renders international collaborations a potential source to understand the dynamics between countries and their advancements.
Analyzing these collaborations can reveal sharing expertise between two countries in different fields, the most well-known institutions of a nation, the overall success of collaborative efforts compared to local ones etc. 
Such analysis were initially performed using statistical metrics \cite{melin1996studying}, but network analysis has later proven much more expressive \cite{wagner2005mapping,gonzalez2008coauthorship}. 
In this exploratory analysis, we aim to examine the collaboration patterns between French and US institutions. Towards this, we capitalize on the Microsoft Academic Graph MAG \cite{sinha2015overview}, the largest open bibliographic dataset that contains detailed information for authors, publications and institutions.
We use the coordinates of the world map to tally affiliations to France or USA. In cases where the coordinates of an affiliation were absent, we used its Wikipedia url and named entity recognition to identify the country of its address in the Wikipedia page. We need to stress that institute names have been volatile (due to University federations created) in the last decade in France, so this is a best effort trial. The results indicate an intensive and increasing scientific production in with , with certain institutions such as Harvard, MIT and CNRS standing out.   

\section{Analysis}

\subsection{Coauthorships of the top French Institutes}
\begin{figure}[h!]
\centering
\includegraphics[scale=.6]{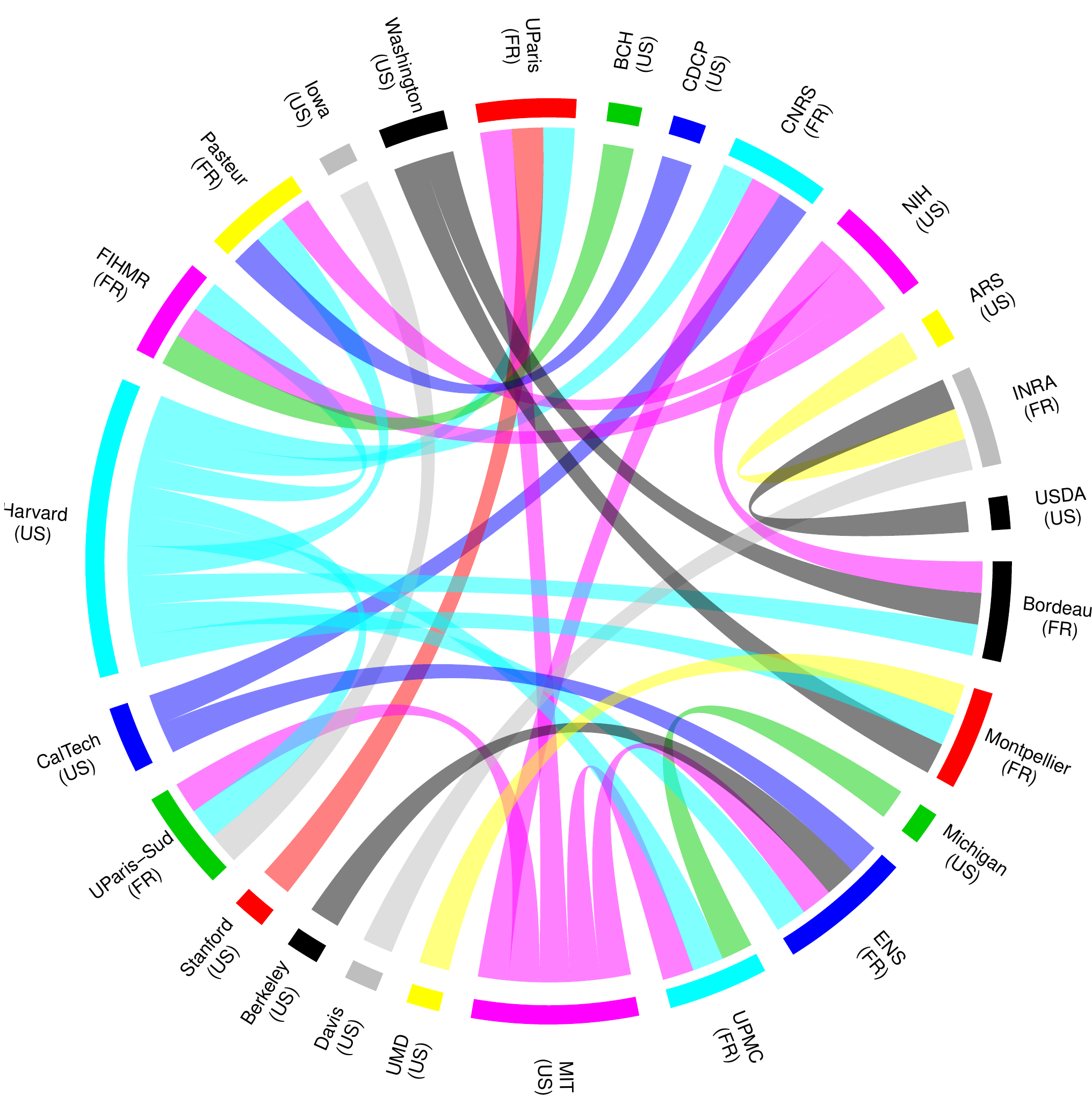}
\caption{Table 1 visualized in a chord plot}
\label{fig:chord}
\end{figure}

We define a collaboration among a French and a US scientific institute if there is at least one paper coauthored by scientists from both these institutes. 
Among the French academic institutes that collaborate with USA we report in Table  \ref{tab:top_french} the 10 most productive (in terms of number of papers) and their most frequent collaborators in the USA. Figure  \ref{fig:chord} visualizes the same information with a chord plot, where the edges are colored based on the US institutes. This shows the US universities collaborating mostly with the aforementioned most productive French institutes. Harvard and MIT stand out with more than 3 collaborations with one of the top French institutes, while the strongest collaboration takes place between CNRS and CalTEch.

\begin{table}[h]
\centering
\begin{tabular}{ |c|c| }
\hline
\textbf{France} & \textbf{USA}\\
\hline
Centre national de la recherche scientifique (CNRS)&\begin{tabular}{@{}c@{}c@{}}California Institute of Technology (CalTech) [2678]\\Massachusetts Institute of Technology(MIT) [2226]\\Harvard University [2106]\end{tabular}\\
  \hline
French Institute of Health and Medical Research (FIHM) &\begin{tabular}{@{}c@{}c@{}}Harvard University [1765] \\National Institutes of Health (NIH) [1484]\\ Boston Children's Hospital (BCH) [1053]\end{tabular}\\
 \hline
University of Paris (UParis)&\begin{tabular}{@{}c@{}c@{}}Harvard University [2061] \\Stanford University [1486] \\Massachusetts Institute of Technology (MIT) [1280]\end{tabular}\\
    \hline
École Normale Supérieure (ENS)&\begin{tabular}{@{}c@{}c@{}}Harvard University [276]\\Massachusetts Institute of Technology (MIT) [252]\\University of California, Berkeley [247]\end{tabular}\\
    \hline
Institut National de la Recherche Agronomique (INRA) &\begin{tabular}{@{}c@{}c@{}}University of California, Davis [255]\\Agricultural Research Service (AGS) [255]\\United States Department of Agriculture (USDA) [244]\end{tabular}\\
  \hline
Pierre-and-Marie-Curie University (UPMC)&\begin{tabular}{@{}c@{}c@{}}Massachusetts Institute of Technology (MIT) [769]\\Harvard University [761]\\University of Michigan [670]\end{tabular}\\
\hline 
 University of Bordeaux&\begin{tabular}{@{}c@{}c@{}}Harvard University [335]\\University of Washington [277]\\National Institutes of Health (NIH) [212]\end{tabular}\\                  
 \hline
 University of Paris-Sud&\begin{tabular}{@{}c@{}c@{}}Massachusetts Institute of Technology (MIT) [1327]\\Harvard University [1305]\\Iowa State University [1093]\end{tabular}\\
 \hline
 University of Montpellier&\begin{tabular}{@{}c@{}c@{}}University of Washington [230]\\Harvard University [219]\\University of Maryland, College Park (UMD) [211]\end{tabular}\\
\hline
Pasteur Institute&\begin{tabular}{@{}c@{}c@{}}National Institutes of Health (NIH) [561]\\Harvard University [421]\\Centers for Disease Control and Prevention (CDCP) [313]\end{tabular}\\
 \hline
\end{tabular}
\caption{Top 10 French institutions (in terms of numbers of papers), with their 3 closest collaborators in US. The number of collaborations in each case is included after the name of the US institute. The names of the institutes are abbreviated for visualization purposes on Fig. \ref{fig:chord}.}
\label{tab:top_french}
\end{table}

\clearpage


\subsection{The timeline of French-US collaborations}

\begin{figure}[h!]
\centering
\includegraphics[scale=.4]{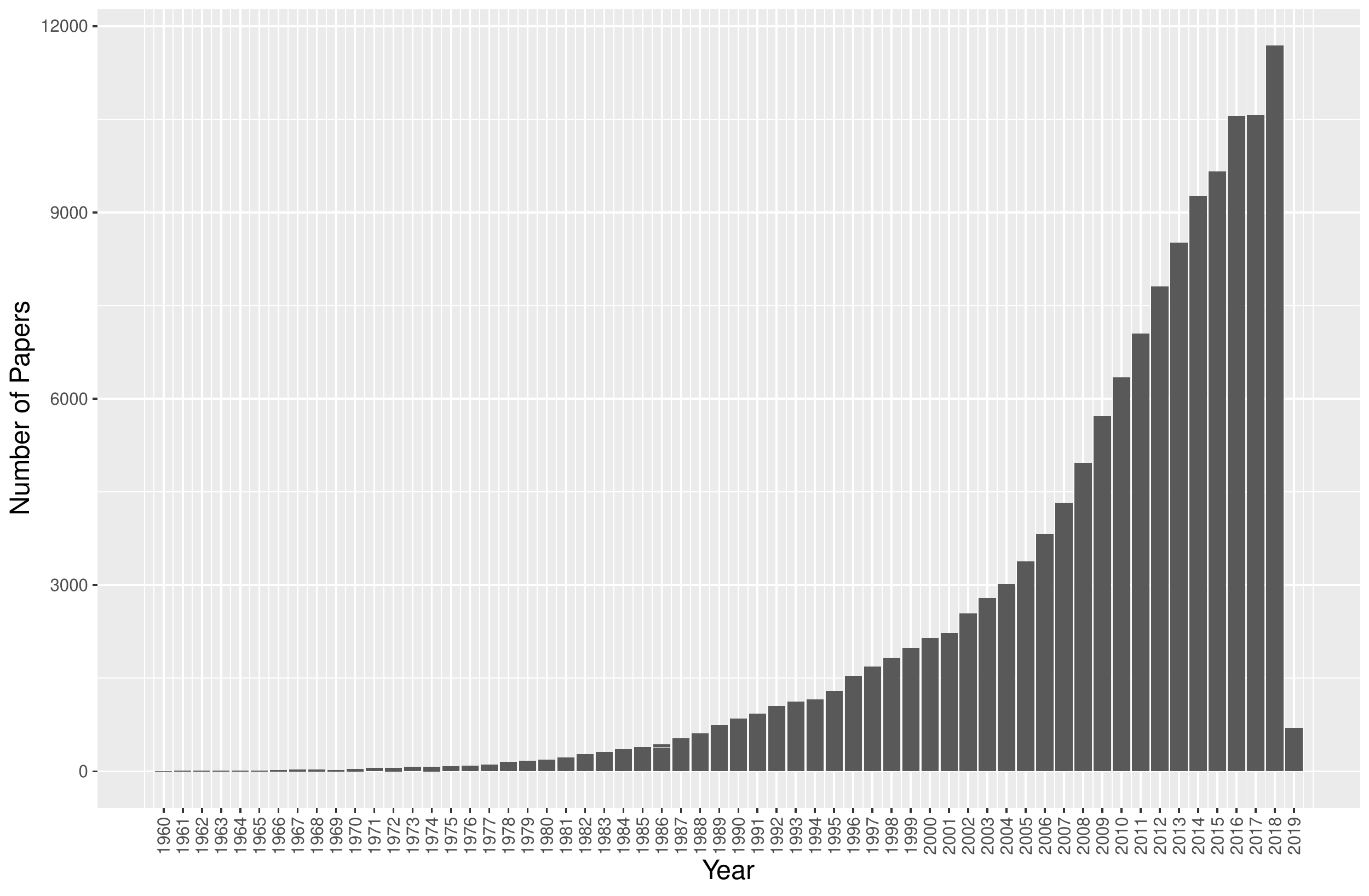}
\caption{FR-US joint papers temporal distribution.}
\label{fig:bar_papers}
\end{figure}

\begin{figure}[h!]
\centering
\includegraphics[scale=.4]{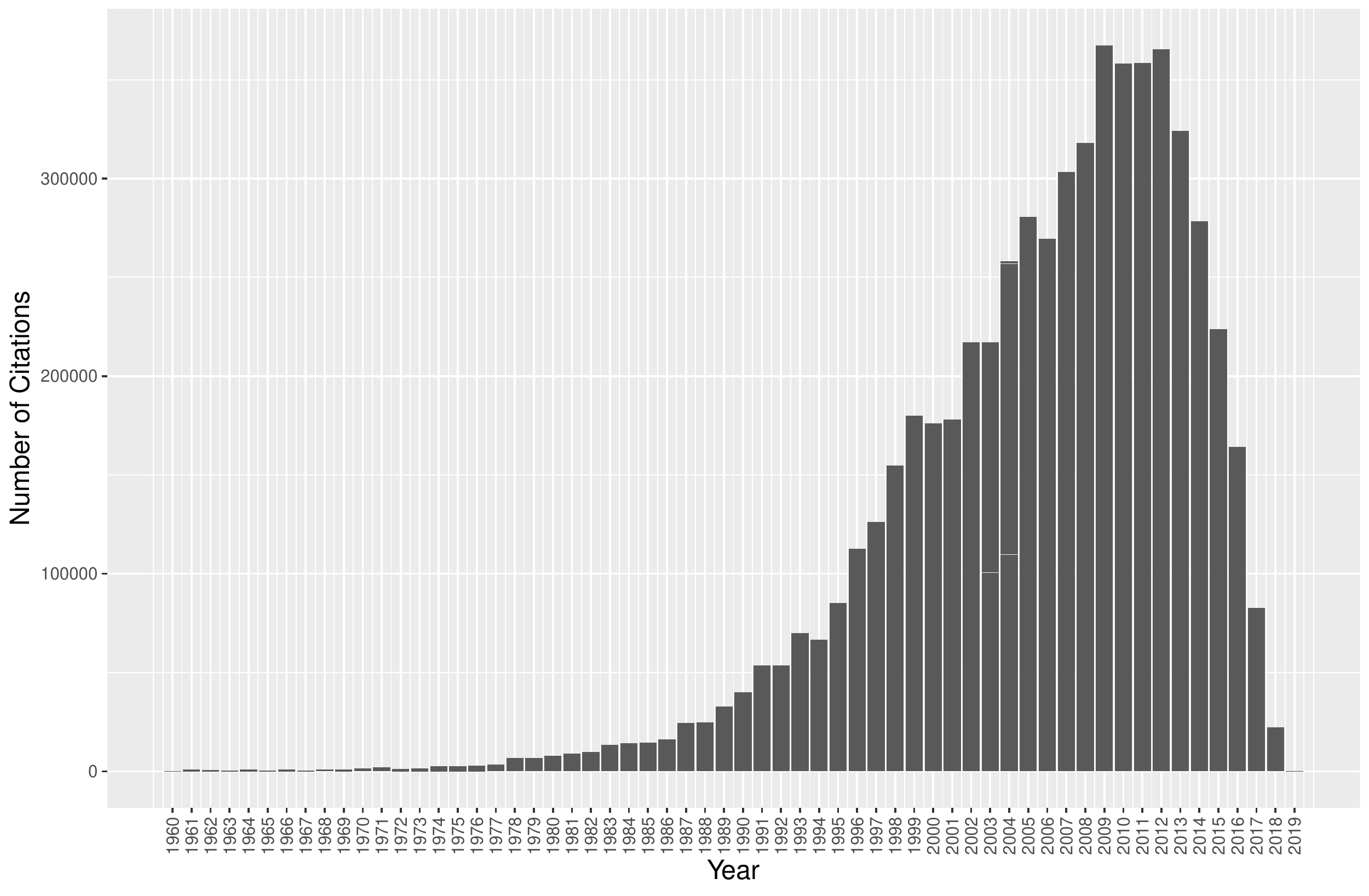}
\caption{FR-US joint papers' citations temporal distribution .}
\label{fig:bar_citations}
\end{figure}

The oldest collaboration between a French and a US institute dates back to 1930, with a joint paper between Ecole Normal Superiere and Cornell University entitled "LA GRAISSE DU SANG ET LA GRAISSE DU LAIT PENDANT LA. LACTATION" \cite{porcher1930graisse} .
Since then, the number of collaborations, as well as their impact (in terms of citations) have increased significantly, as one can see in Figures \ref{fig:bar_papers} and  \ref{fig:bar_citations}.
There is an almost exponential increase in the number of papers produced by joint collaborations. The same applies for the impact of these works, which increases especially until 2010.Naturally, the citations count in recent years is diminished, as young papers have fewer citations and they need time to get cited. There are also some monumental years, such as 2012, where the inclusion of the publications on the discovery of the Higgs Boson particle e.g. \cite{aad2012observation,bezrukov2012higgs} has produced a massive number of citations.

\begin{center}
\begin{figure}[h]
\includegraphics[scale=.9]{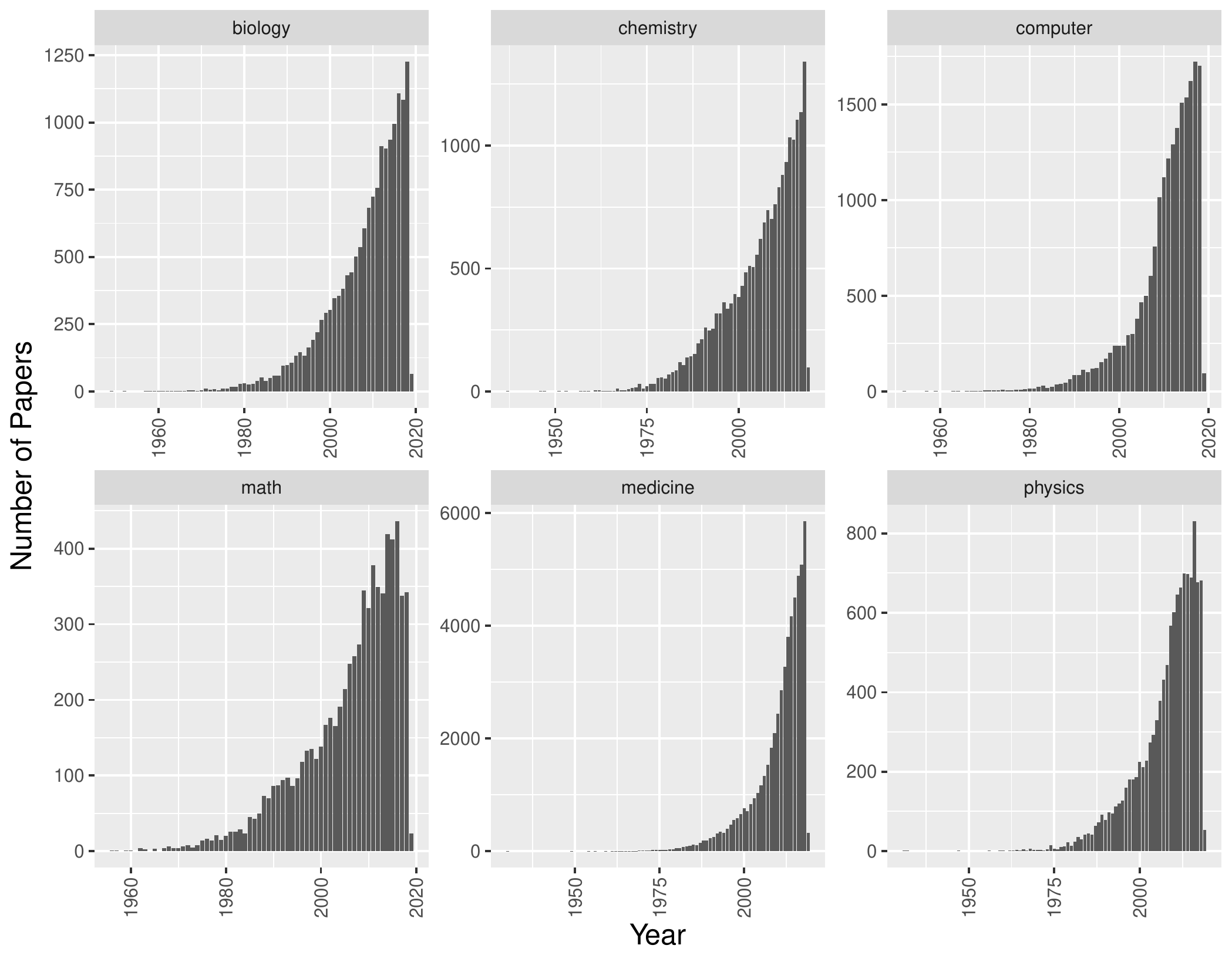}
\caption{FR-US joint papers temporal distribution per prominent scientific area.}
\label{fig:bar_field_papers}
\end{figure}
\end{center}

\begin{center}
\begin{figure}[h]
\includegraphics[scale=.9]{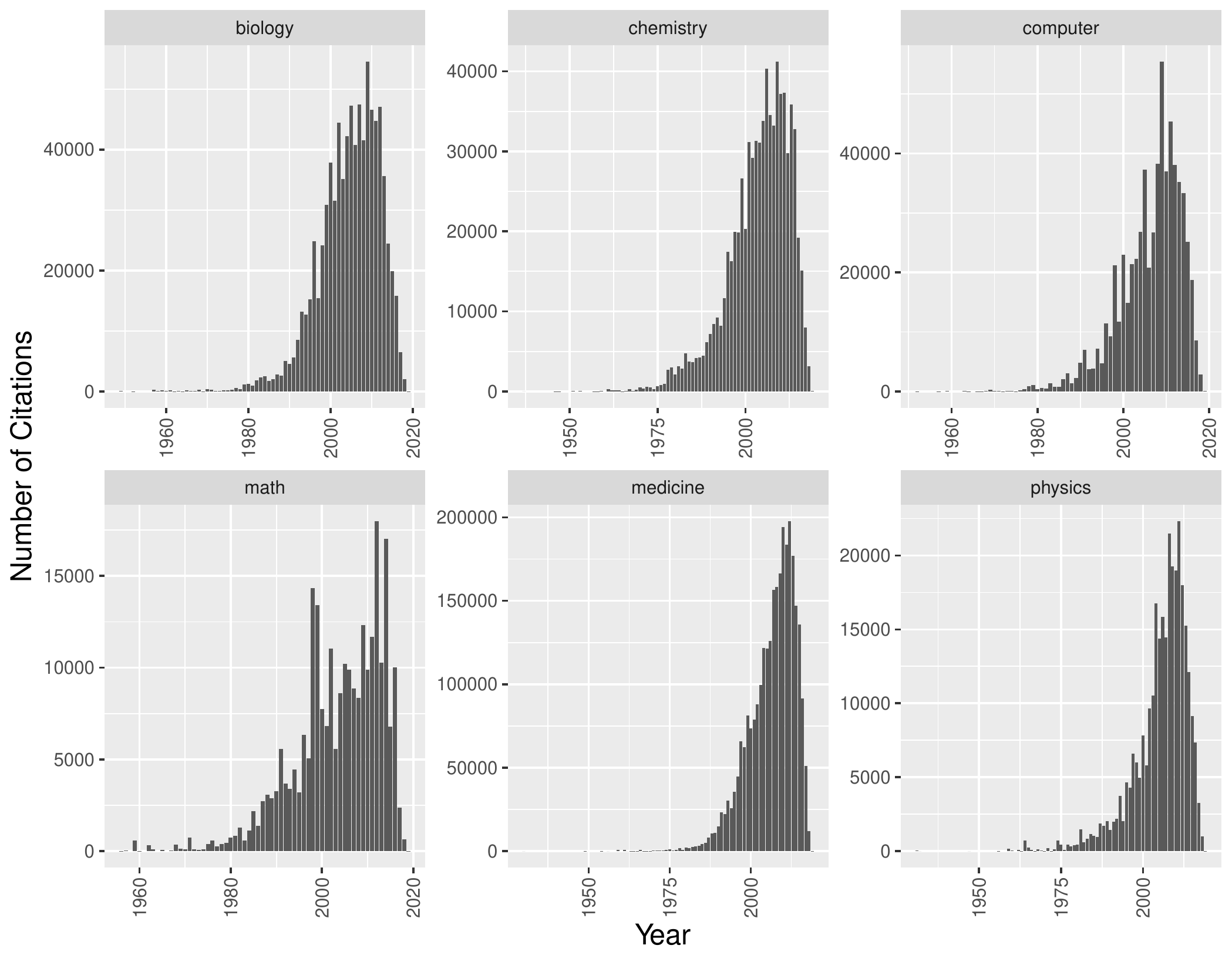}
\caption{FR-US joint papers \textit{citations} temporal distribution per prominent scientific area.}
\label{fig:bar_field_citations}
\end{figure}
\end{center}

Distinguishing between different fields, we can see in Fig. \ref{fig:bar_field_papers} how the number of collaborations have evolved through the years in different disciplines, as well as their success in Fig. \ref{fig:bar_field_citations}. Bare in mind that a paper might belong to more than one fields.
\begin{itemize}
    \item We see that the majority of works are comprised of medical studies, which is quite common in academia.
    \item The second field is computer science and the third biology. Especially in computer science, there is a steep increase around 2010.
    \item The citations do not follow the same pattern, as the top cited domain is  medicine, followed by biology and and computer science. This is due to  the known differences in citation patterns among computer science and biology \cite{patience2017citation}.
    \item Mathematics is the least active field in this context since publications in this area are relatively rare. Still there are some spikes of citations through the years because of some important papers.
\end{itemize}

\clearpage

\subsubsection{Top Collaborations}
\begin{figure}[h!]
\centering
\includegraphics[scale=.4]{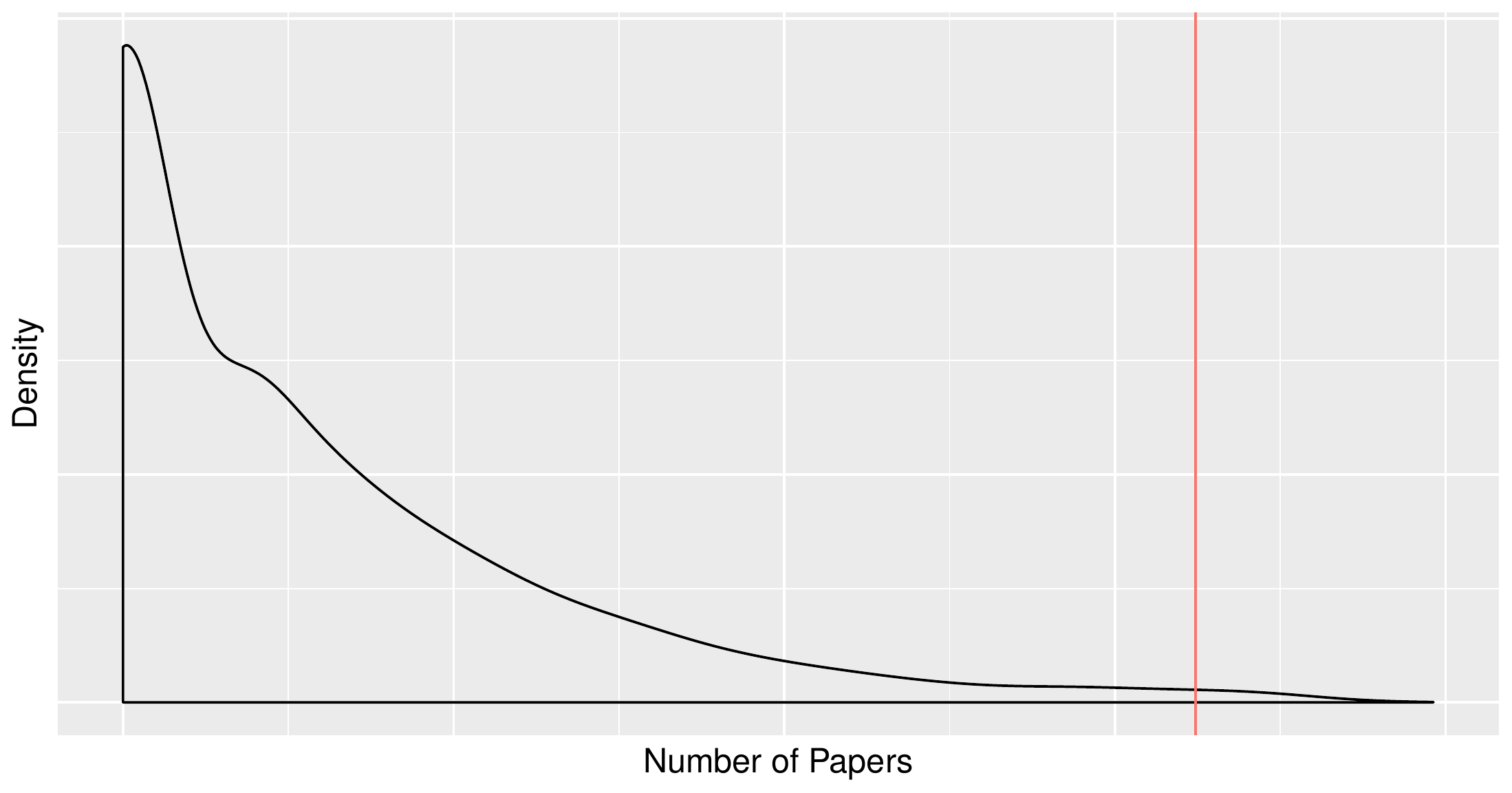}
\caption{Density of the number of collaborations with the number of papers, and the threshold to define the top collaborations.}
\label{fig:density}
\end{figure}

Let the strength of the collaboration $c$ be denoted as $P(c)$, refering to the number of such papers. We rank all collaborations based on strength and define the most prominant the ones who are above the average number of papers in all collaborations and six standard deviations of the distribution i.e. $\{c | P(c)\geq mean({\{P(c')\}_{c'\in C}})+6sd(\{P(c')\}_{c'\in C})\}$, where $C$ is the set of all collaborations. The threshold and the density are visualized in figure \ref{fig:density}. The number of collaborations that are above this threshold is 492.

\begin{figure}[h!]
\centering
\includegraphics[scale=.9]{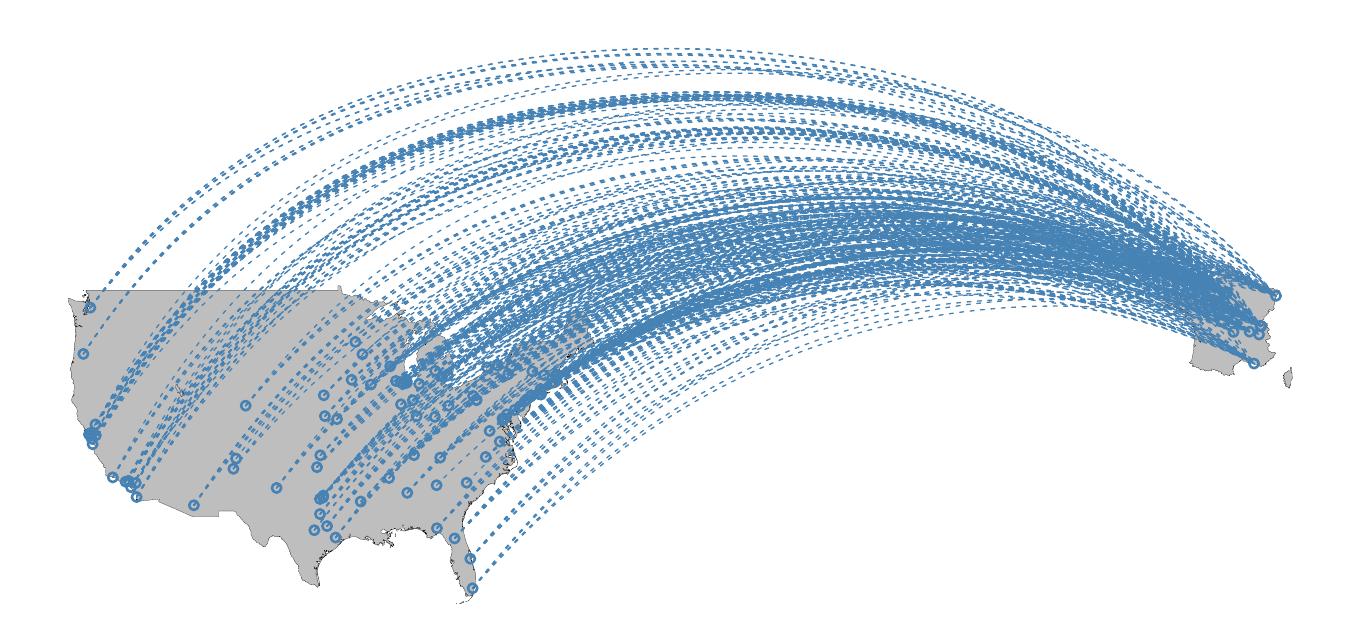}
\caption{USA-France top collaborations.}
\label{fig:collab_big}
\end{figure}

\begin{figure}[h!]
\centering
\includegraphics[scale=.4]{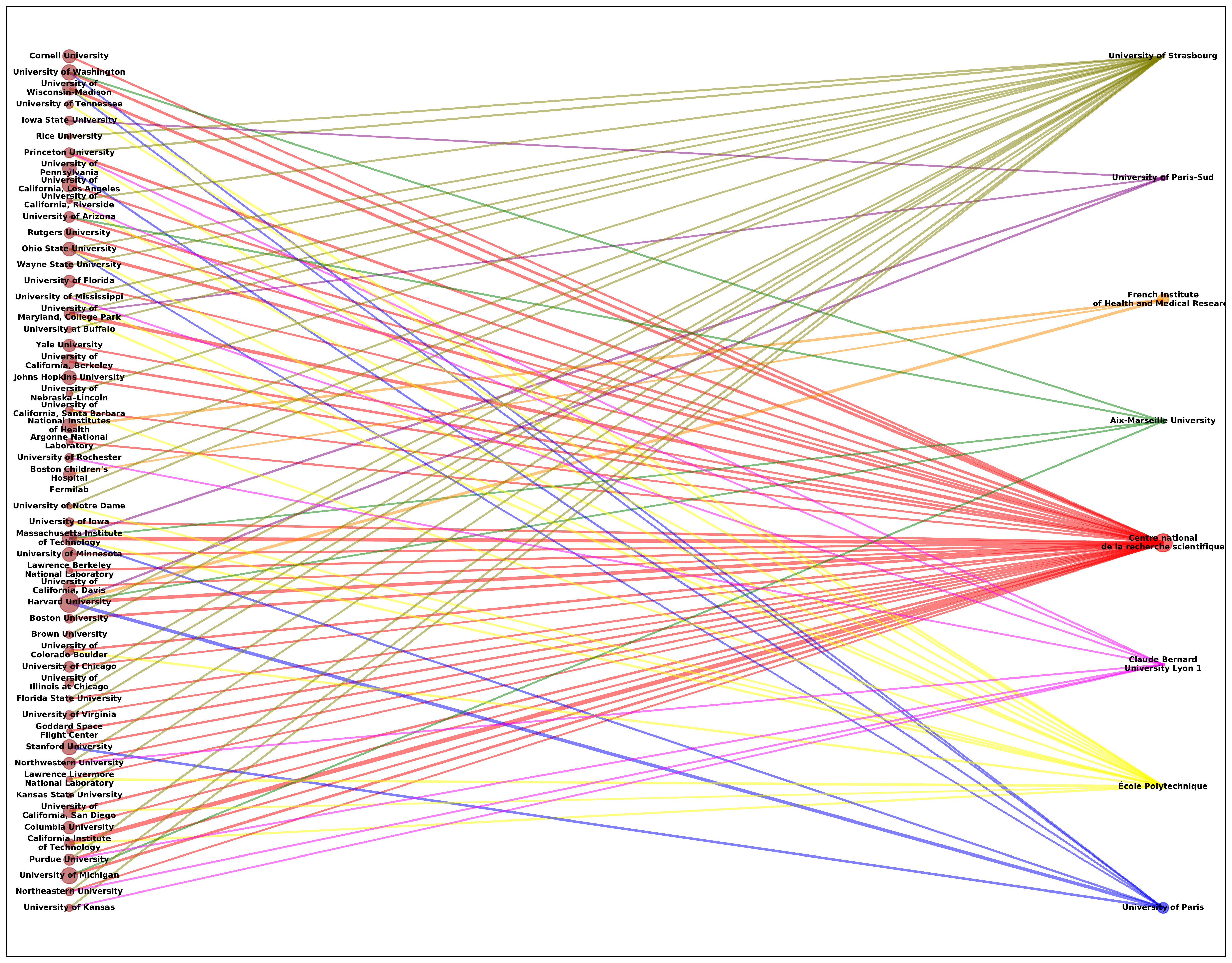}
\caption{USA-France top 100 collaborations in terms of number of joint papers. Left is USA,  right is France. The size of the nodes is proportional to their number of papers, the edge width is proportional to the collaborations' strength and their color is different for each French institution.}
\label{fig:collab_small}
\end{figure}

These are visualized in the map plot \ref{fig:collab_big}. Although we can get an idea of the overall cities that collaborate mostly with each other, it is still a very perplex image to make sense of. Thus we reduce it even more, by taking the top 100 collaborations, and making a weighted bipartite plot in \ref{fig:collab_small}.

Few observations derived from a first glance on this network:
\begin{itemize}
    \item CNRS is the French institute with the most collaborations, followed by Ecole Polytechnique and University of Strasbourg.
    \item Some strong collaborations that stand out are CNRS with CalTech and UMD and University of Paris with Harvard.
    \item The French Institute of Health and Medical Research has few collaborations, but with two US institutes well known for their achievements in medicine, Harvard and NIH.
    \item Institutes performing research on fields like physics, biology or chemistry tend to have more connections than the ones focusing on language studies or the ones performing solely medical research. This might indicate that STEM projects applied or related to medicine exhibit international collaborations, while purely medical studies have a local collaboration pattern. 
    \item The most productive French Institute is clearly CNRS, followed by University of Paris, while for the USA it is Harvard and University of Michigan.
\end{itemize}
\clearpage

\subsubsection{Top 10 collaborations}
The top 10 USA-France collaborations in terms of absolute number of joint papers can be seen in table \ref{tab:top_edges}
   
\begin{table}[h]
\centering
\begin{tabular}{ |c|c| }
\hline
\textbf{USA}&\textbf{France}\\
\hline
California Institute of Technology    &  Centre national de la recherche scientifique (CNRS)  2678\\ 
 \hline
Massachusetts Institute of Technology & Centre national de la recherche scientifique  (CNRS) 2226\\  
 \hline
  Harvard University   & Centre national de la recherche scientifique (CNRS)  2106\\
 \hline
University of Maryland, College Park    & Centre national de la recherche scientifique  (CNRS) 2092  \\
 \hline
Harvard University   & University of Paris   2061\\
\hline
Ohio State University   &Centre national de la recherche scientifique (CNRS)  1840  \\
\hline
 Harvard University   &French Institute of Health and Medical Research   1765   \\
\hline
University of Wisconsin-Madison      &   Centre national de la recherche scientifique (CNRS)  1742\\
\hline
Princeton University &   Centre national de la recherche scientifique (CNRS)  1638\\
\hline
University of Michigan &   Centre national de la recherche scientifique (CNRS)  1596\\
\hline
\end{tabular}
\caption{Top USA-France collaborations in terms of number of papers.}
\label{tab:top_edges}
\end{table}

For each of the 10 collaborations we report the topical distribution in the figures below.
Some observations:
\begin{itemize}
    \item The collaboration between CalTech and CNRS focuses mainly in math,chemistry and physics, since both are widely known for their excellence in these fields.
    \item The joint works of CNRS with MIT and Harvard are predominantly related to medicine. In contrast, the secondary fields in MIT are computer science, chemistry, physics and math, while for Harvard biology is second and the rest follow. This makes sense because, as mentioned above, MIT is focusing more on science and technology while Harvard has a broader scope.
    \item University of Maryland (UMD) and CNRS seem to collaborate in a variety of disciplines, with physics being the  dominant. 
    \item It is clear the main joint papers between Harvard and University of Paris refer to medical studies. The same applied to the French Institute of Health and Medical Research, which overall might refer to joint works between these three institutes.
    \item CNRS has also collaborated extensively with Ohio State, Wisconsin-Madison and Michigan, especially in medicine. 
\end{itemize}

\begin{figure}[h!]
\centering
\begin{minipage}{.5\textwidth}
  \centering
  \includegraphics[scale=.3]{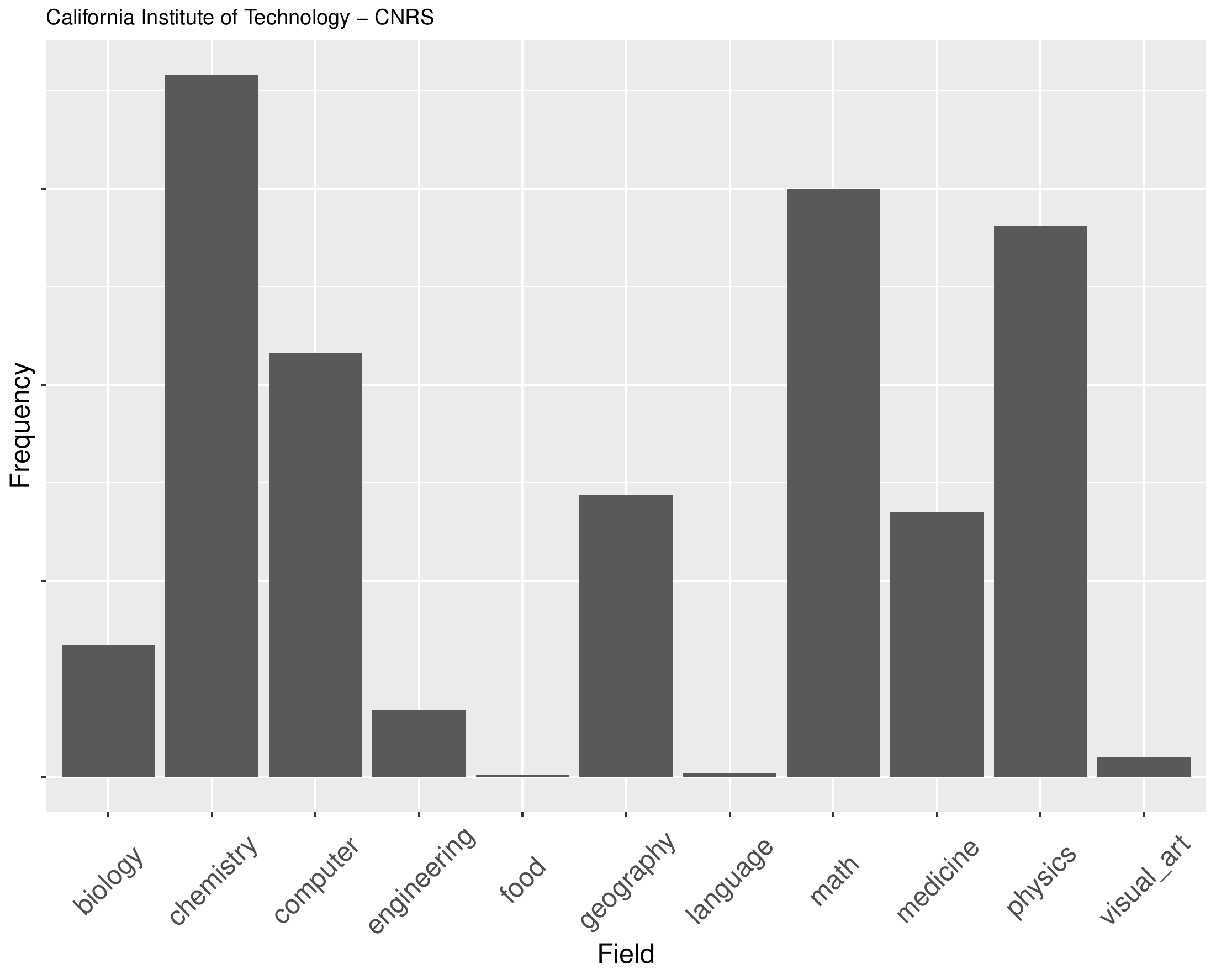}
\end{minipage}%
\begin{minipage}{.5\textwidth}
  \centering
  \includegraphics[scale=.3]{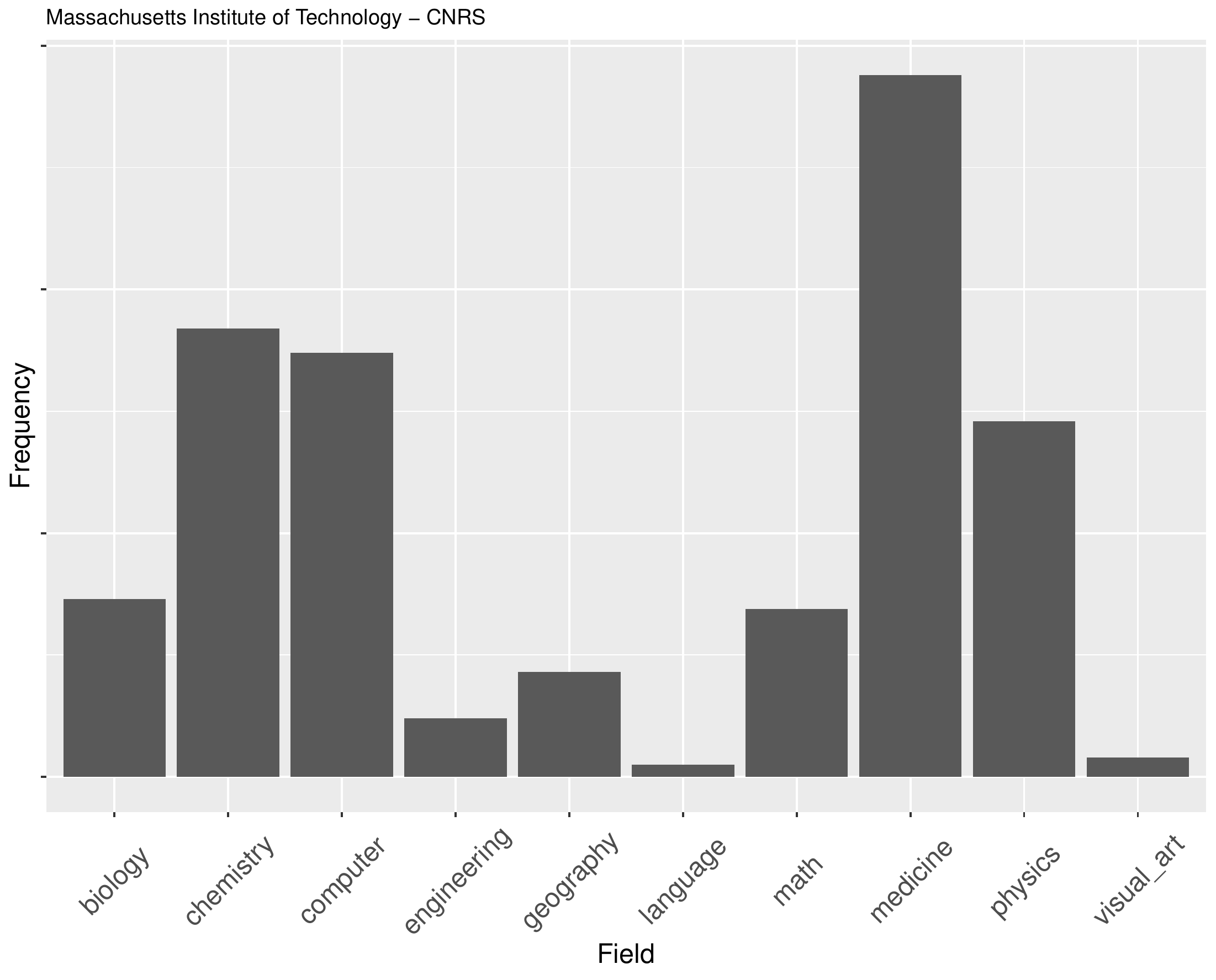}
\end{minipage}
\end{figure}

\begin{figure}[h!]
\centering
\begin{minipage}{.5\textwidth}
  \centering
  \includegraphics[scale=.3]{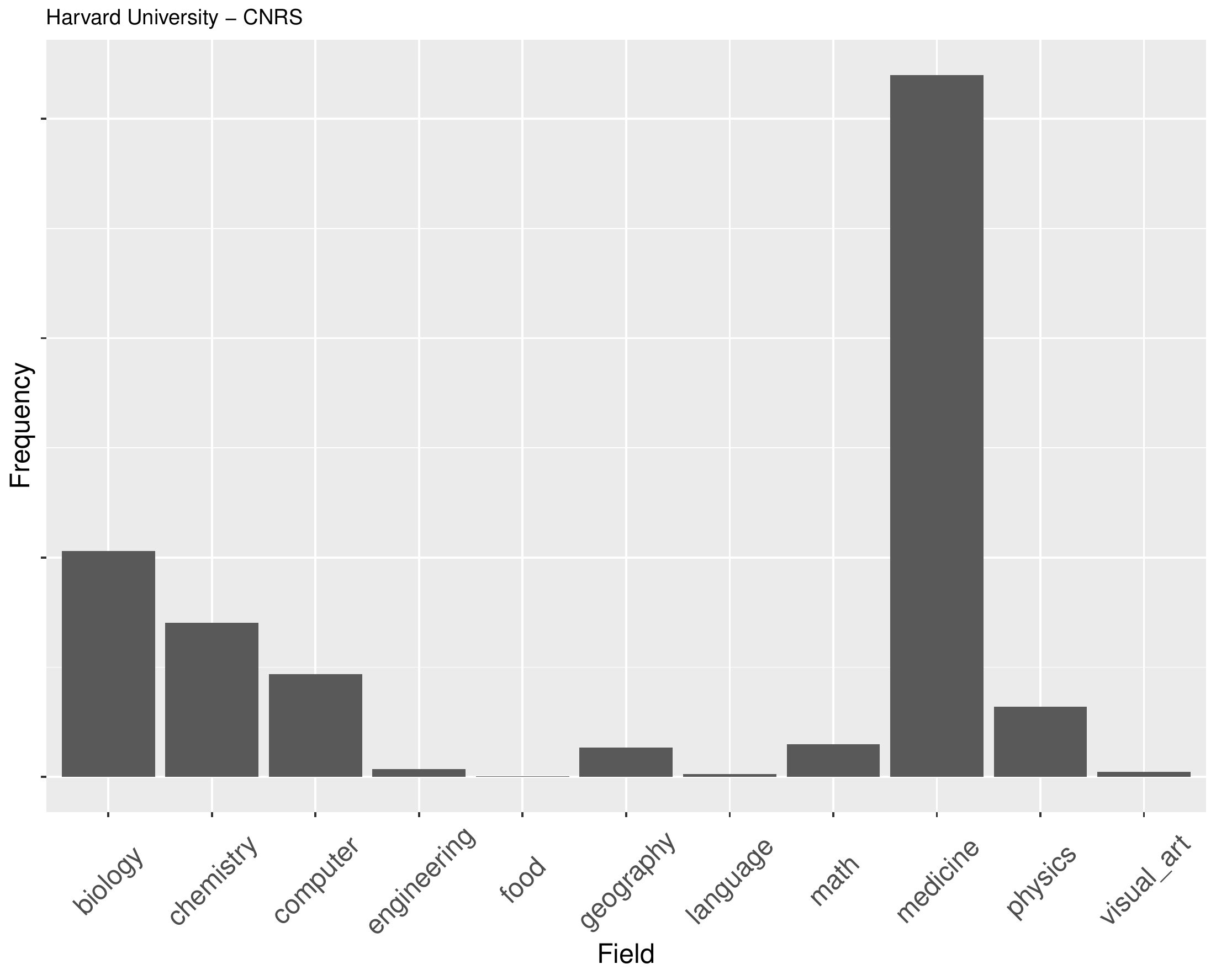}
\end{minipage}%
\begin{minipage}{.5\textwidth}
  \centering
  \includegraphics[scale=.3]{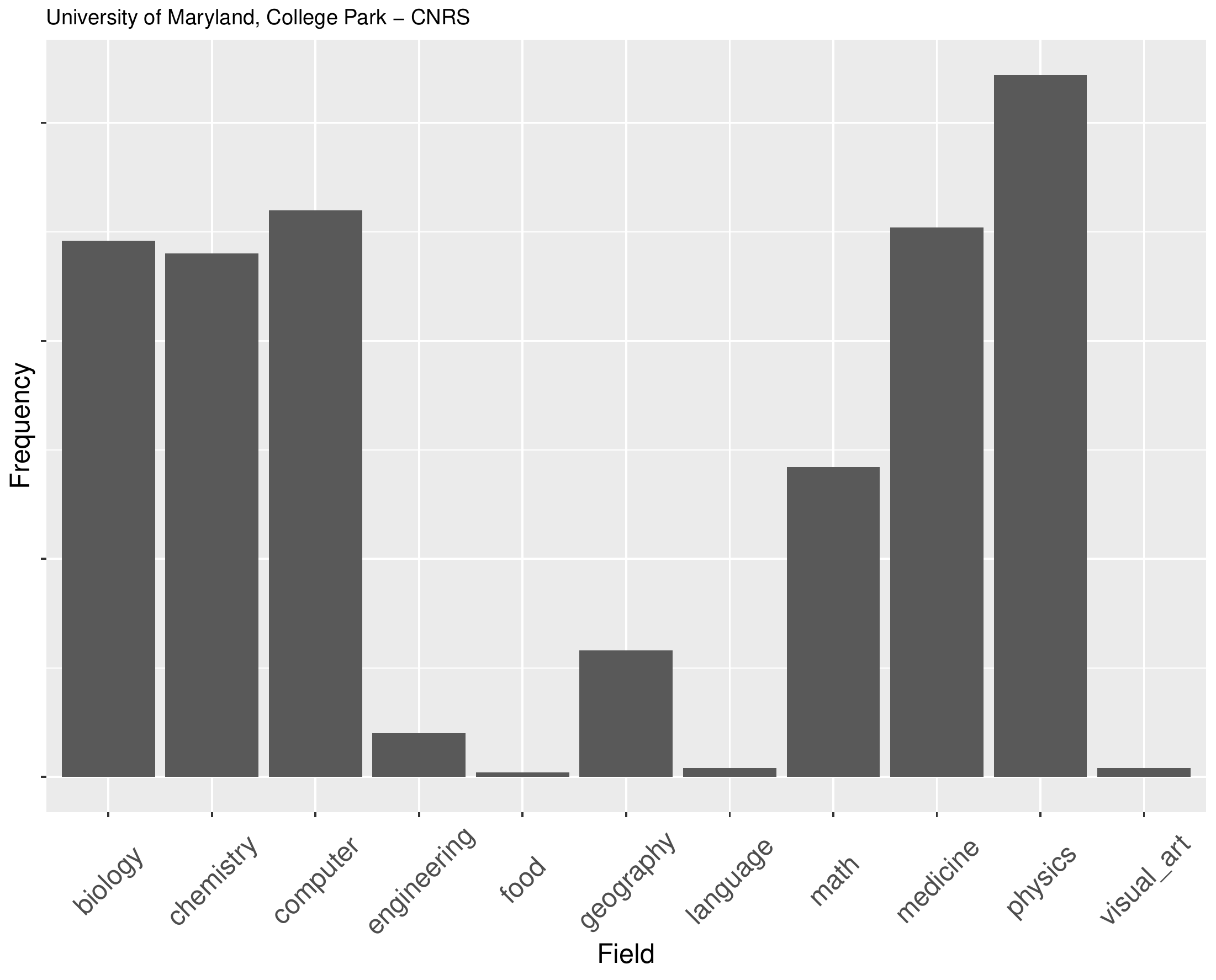}
\end{minipage}
\end{figure}

\begin{figure}[h!]
\centering
\begin{minipage}{.5\textwidth}
  \centering
  \includegraphics[scale=.3]{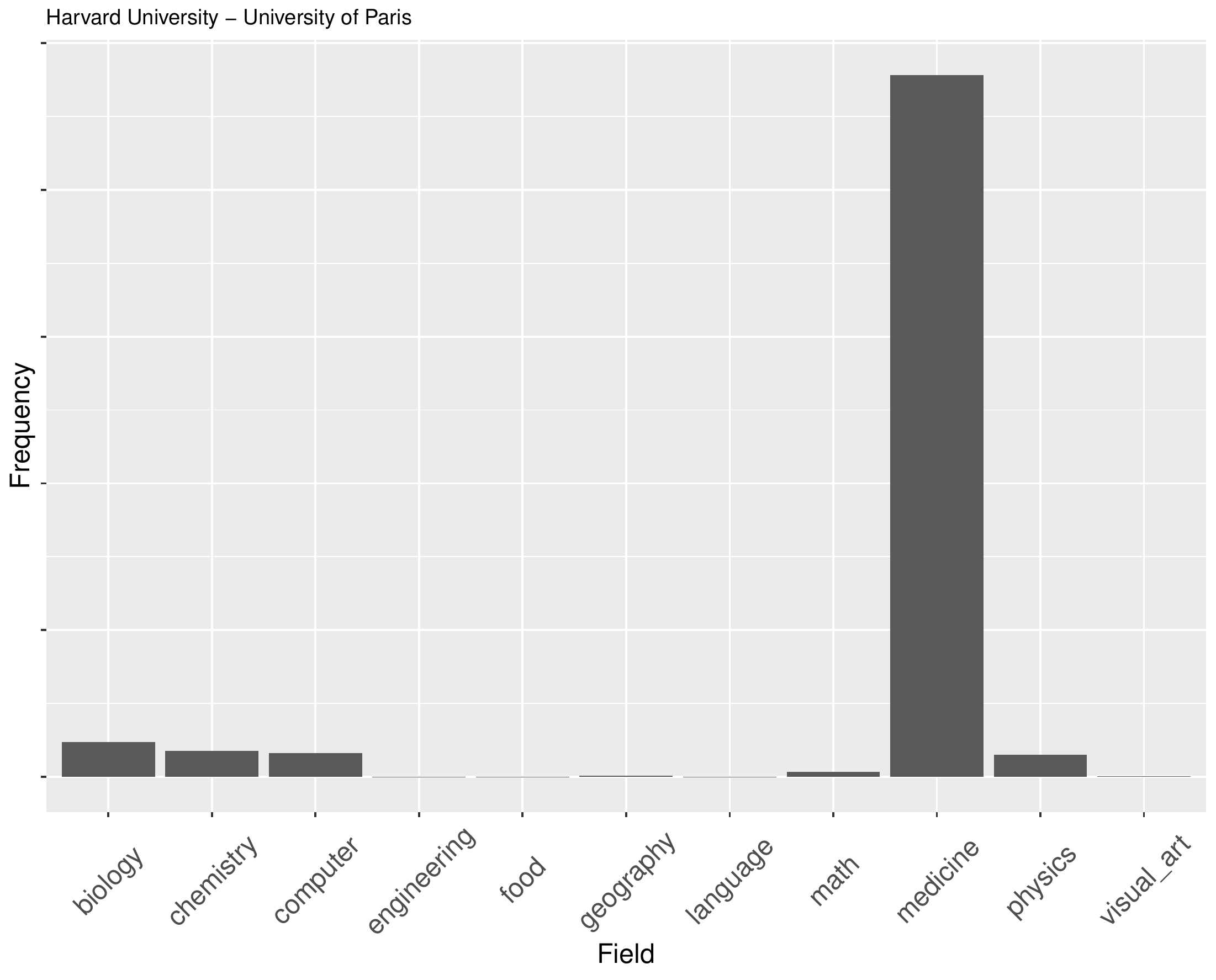}
\end{minipage}%
\begin{minipage}{.5\textwidth}
  \centering
  \includegraphics[scale=.3]{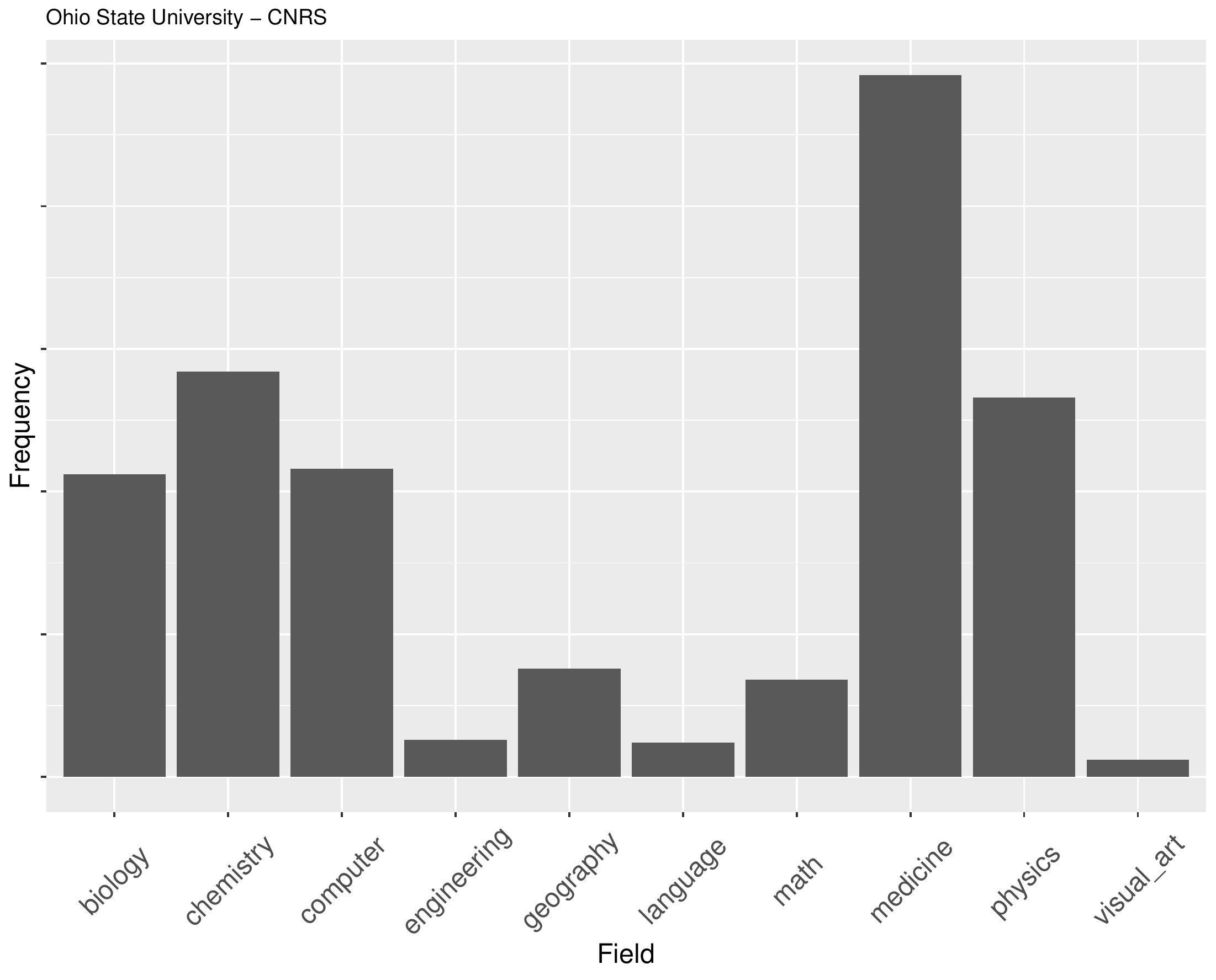}
\end{minipage}
\end{figure}

\begin{figure}[h!]
\centering
\begin{minipage}{.5\textwidth}
  \centering
  \includegraphics[scale=.3]{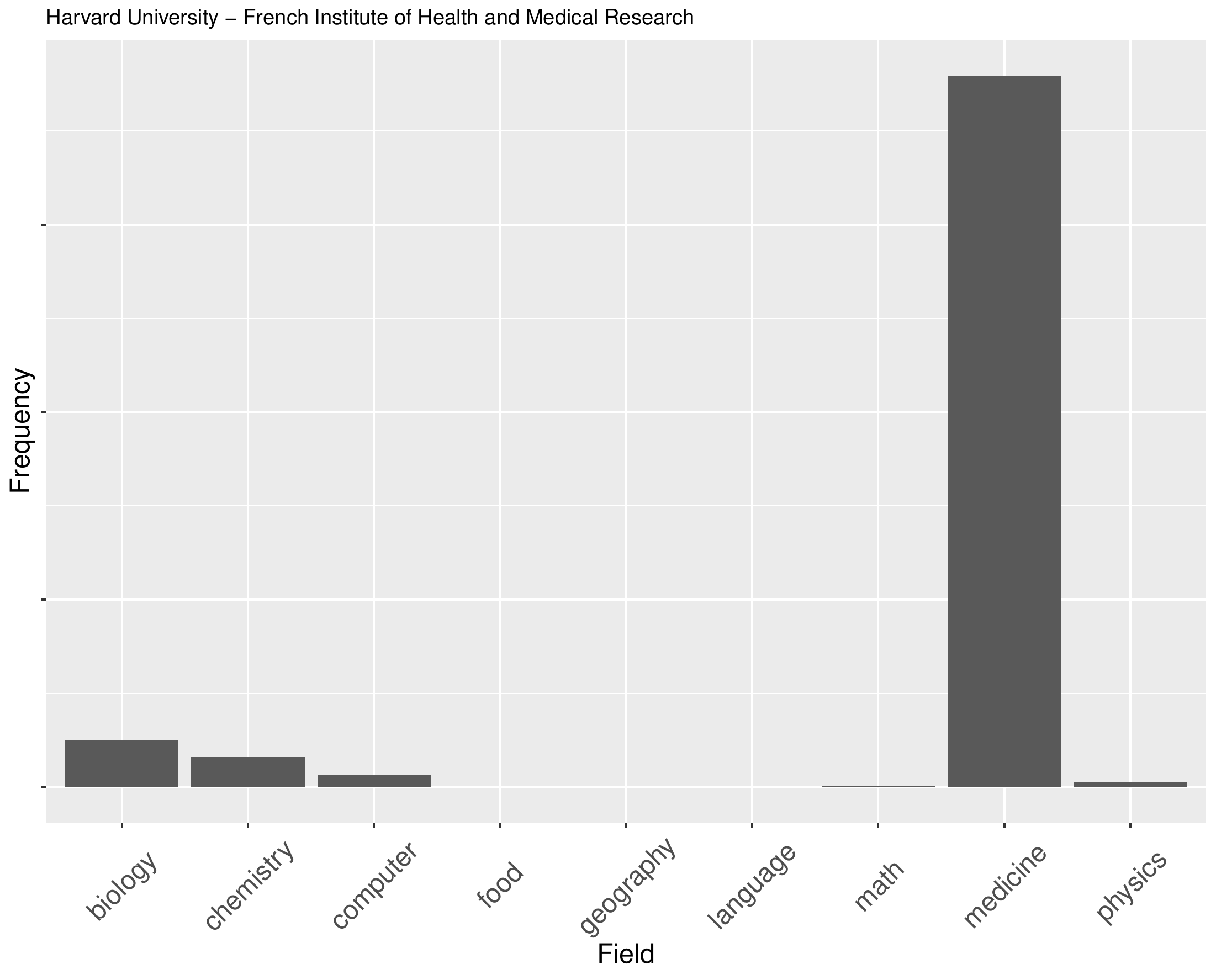}
\end{minipage}%
\begin{minipage}{.5\textwidth}
  \centering
  \includegraphics[scale=.3]{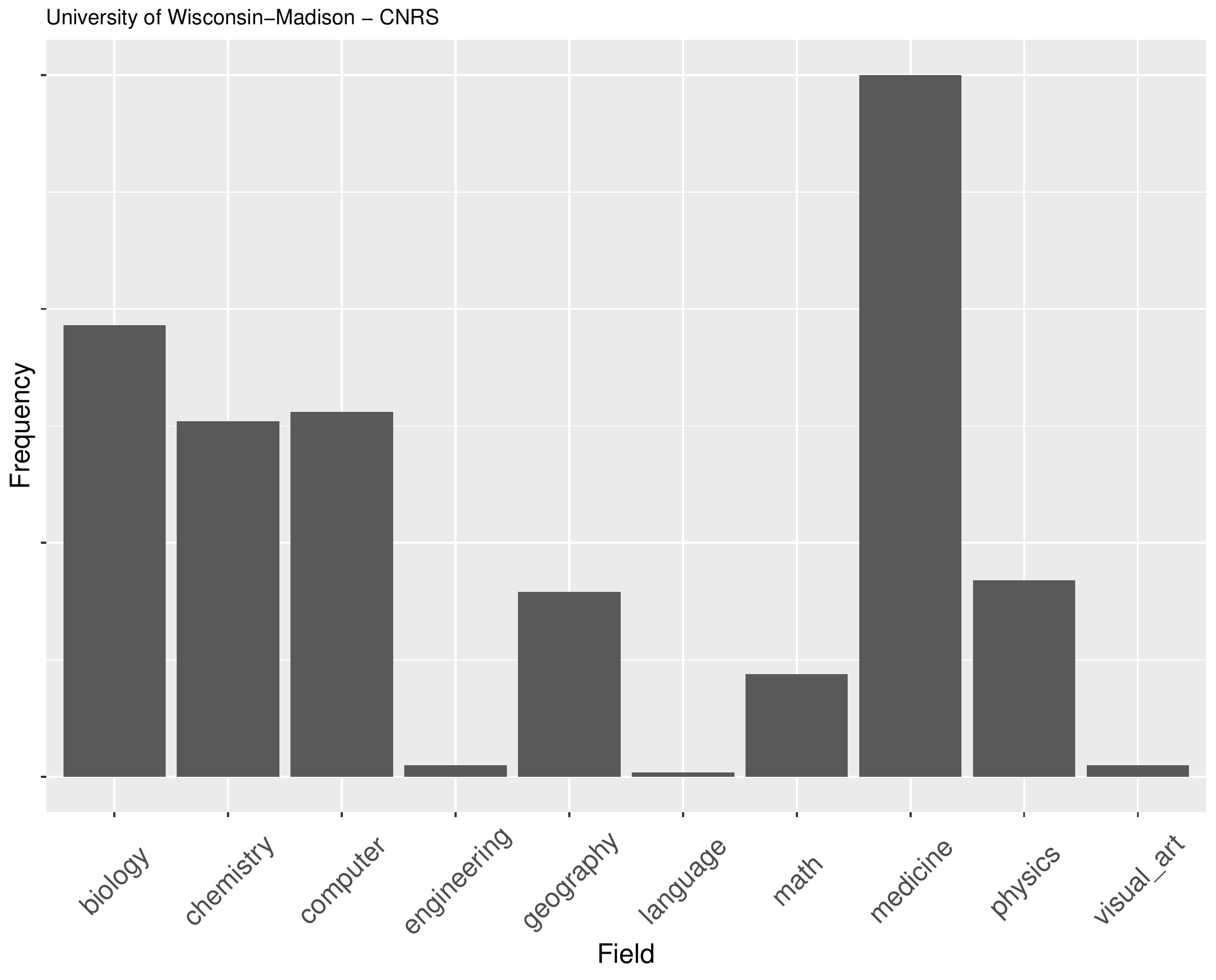}
\end{minipage}
\end{figure}

\begin{figure}[h!]
\centering
\begin{minipage}{.5\textwidth}
  \centering
  \includegraphics[scale=.3]{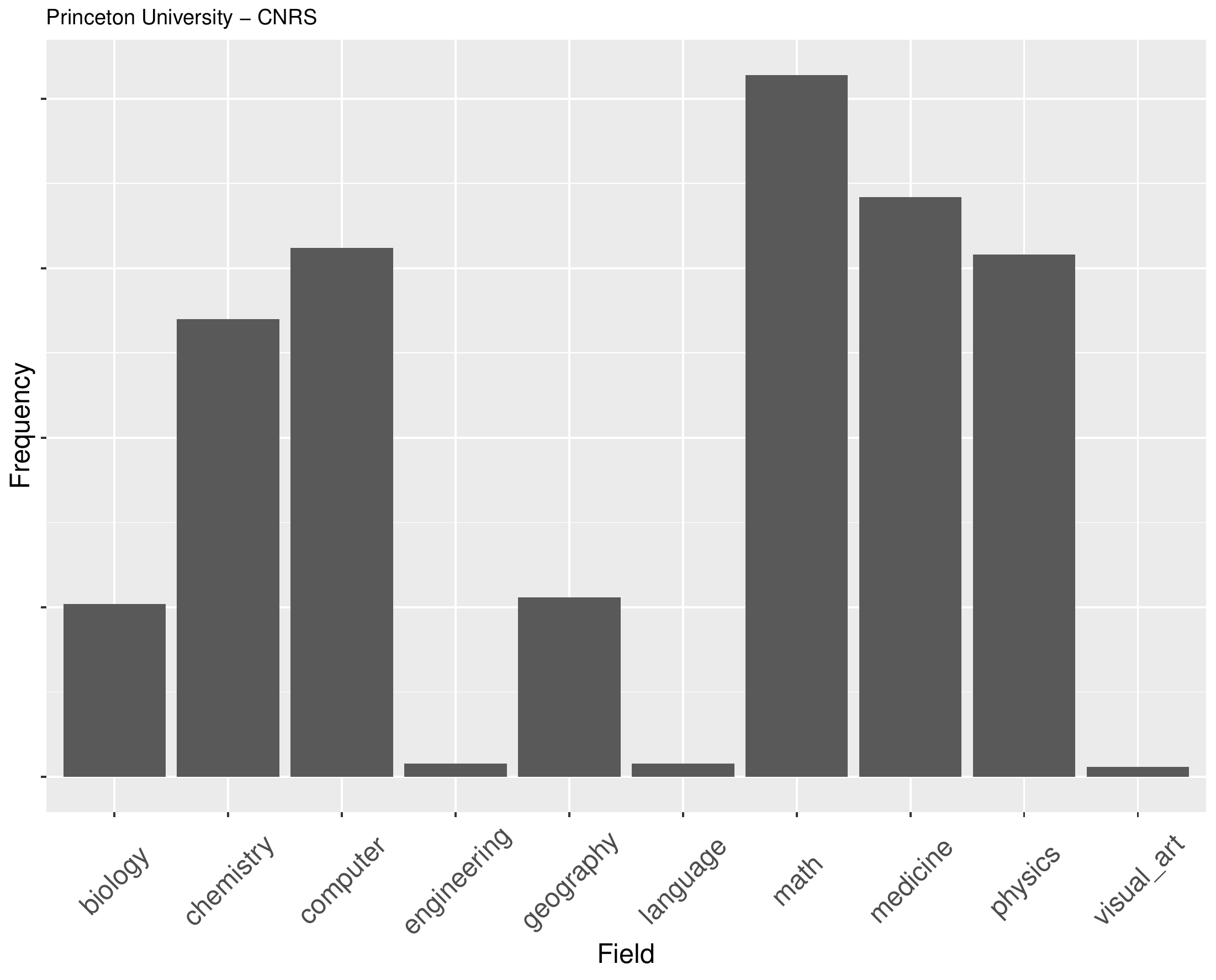}
\end{minipage}%
\begin{minipage}{.5\textwidth}
  \centering
  \includegraphics[scale=.3]{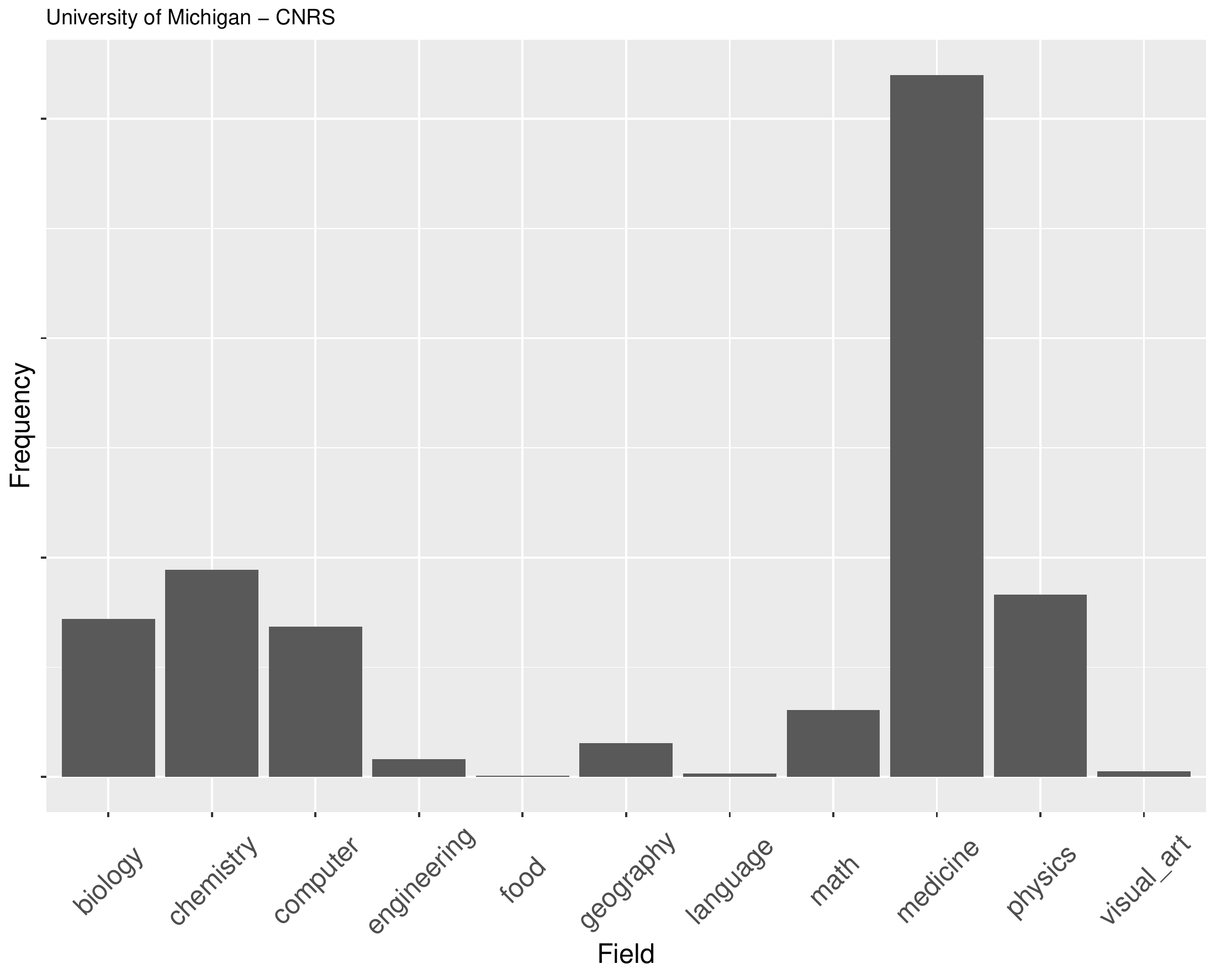}
\end{minipage}
\end{figure}

\clearpage

\section{Future Work}
These initial results indicate there is an intensive and increasing scientific production in terms of joint papers and resulting citations. Overall, several new hypotheses can be tested:
\begin{itemize}
    \item The use of influence and success metrics to identify the most crucial authors \cite{giatsidis2019rooted,panagopoulos2019scientometrics,panagopoulos2020multi}, cliques or laboratories that guide the course of the majority of the collaborations in a direct or indirect manner.
\item Prediction of new collaborations based on the fields the institutes belong to or the venues they publish at.
\item Prediction of institutes that will increase their long-term impact based on their current position and activity in the network \cite{panagopoulos2017detecting}.
    \item Use of complementary datasets such as OpenAIRE \cite{manghi2010infrastructure} to include information about funded projects.
    
\end{itemize}

\section{Acknowledgements}
This analysis was performed after the request of Dr. Yves Frenot, Dr. Jean-Baptiste Bordes and Maxime Benallaoua from the Office for Science and Technology of the Embassy of France in the United-States, with whom we collaborated to set the hypotheses examined.
\bibliographystyle{spm}
\bibliography{bib}

\end{document}